# Experimental Characterization of the Energetics of Low-temperature Surface Reactions

Thomas K. Henning[1] and Serge A. Krasnokutski[2*]

[1]Max Planck Institute for Astronomy, Königstuhl 17, 69117 Heidelberg, Germany; [2]Laboratory Astrophysics Group of the Max Planck Institute for Astronomy at the University of Jena, Helmholtzweg 3, 07743 Jena, Germany. *Email: sergiy.krasnokutskiy@uni-jena.de.

**Astrochemical surface reactions are thought to be responsible for the formation of complex organic molecules, which are of potential importance for the origin of life. In a situation, when the chemical composition of dust surfaces is not precisely known, the fundamental knowledge concerning such reactions gains significance. We describe an experimental technique, which can be used to measure the energy released in reactions of a single pair of reactants. These data can be directly compared with the results of quantum chemical computations leading to unequivocal conclusions regarding the reaction pathways and the presence of energy barriers. It allows for predicting the outcomes of astrochemical surface reactions with higher accuracy compared to that achieved based on gas-phase studies. However, for the highest accuracy, some understanding of the catalytic influence of specific surfaces on the reactions is required. The new method was applied to study the reactions of C atoms with $H_2$, $O_2$, and $C_2H_2$. The formation of HCH, CO + O, and triplet cyclic-$C_3H_2$ products has been revealed, correspondingly.**

   Surface reactions play a key role in molecule formation in astrophysical environments with the most abundant molecule, $H_2$, being a prominent example[1]. Associative reactions rarely operate under gas-phase conditions. However, they are important in increasing the molecular size and complexity in astrophysical environments. Associative reactions are only allowed when a third body for energy and momentum transfer is present. The third body can be any kind of species or medium. In particular, it could be the surface of cosmic grains. This is the reason that surface chemistry is often considered to play a key role in the formation of saturated complex organic molecules (COMs)[2].

The COMs are detected in the gas phase[2], and it is expected that even larger and consequently more refractory molecules are present as a "sticky" solid component on the surface of refractory silicate and carbonaceous dust. A wealth of COMs has been discovered by (sub)millimeter spectroscopy[3] of molecular cloud cores and protoplanetary disks, the Rosetta mission to comet 67P/Churyumov–Gerasimenko[4], and by the in-situ analysis of primitive meteorites – the carbonaceous chondrites[5]. The continuous delivery of such pre-biotic organic molecules, including a variety of nucleobases, amino acids, and sugars, to early Earth or Earth-like planets in other planetary systems, may have triggered the formation of life. At the appropriate planetary conditions, these molecules can polymerize and are potential precursors of the RNA world[6,7].

In contrast to the fundamental importance of surface chemistry for the formation of COMs, there is a general lack of quantitative experimental data on the reactions relevant to astrochemistry. In the various astrochemical reaction databases, the number of included surface reactions is almost one order of magnitude lower compared to the number of gas-phase reactions[8]. Moreover, the pathways of these surface reactions are often first guesses only, mostly based on gas-phase reactions which could result in the application of wrong reaction channels. This situation affects even the most important astrochemical reactions between abundant species. We note as an example that the surface reaction $C + H_2 \rightarrow HCH$ has been recently found to be barrierless[9] in contrast to the energy barrier of 2500 K considered in the database[8].

In this study we introduce a completely new and innovative experimental technique designed to characterize the landscape of low-temperature chemical reactions. The technique allows us to directly measure the energy released in reactions of a single pair of reactants. In this way, the method exploits the same approach as was used in the pioneering studies by Yuan Lee to characterize gas-phase reactions by crossed molecular beam techniques[10, 11]. So far, the calorimetric techniques available to study surface reactions, like single crystal adsorption calorimetry or isothermal titration calorimetry, provide the enthalpy of the reactions of molecules with a surface[12]. To the best of our knowledge, there is still no universal technique available that allows us to precisely measure the amount of energy released in surface reactions of a single pair of reactants.

In our method, the reactions are studied inside liquid helium, which has a negligible influence on the reaction energy levels. This is evidenced by minor shifts of electronic transitions caused by liquid helium[13]. Tiny helium nanodroplets are used as nanocalorimeters measuring the amount of energy released in the reaction of a single pair of reactants.

We employed the He droplet isolation technique pioneered by Scoles and Toennies[13-15] and developed a procedure that allows us to measure the amount of energy released in the reaction of a single pair of reactants inside the droplets. In this method, we used the advantage of the extremely small size of the He droplets and the fact that evaporation of each He atom removes about 7.2 K of equivalent energy[13]. Therefore, the evaluation of a change in He droplet sizes after reactions provides an estimate of the released amount of energy. This method was first used to evaluate the energy released during cluster growth[16] and later to detect the energy released during a chemical reaction[9, 17-20].

However, so far the method was inaccurate and could not provide the exact amount of energy released in the reaction of a single pair of reactants. A more precise method[21,22] is only available for the evaluation of the reaction energy between identical species picked up by He droplets from the gas phase at the same location. However, even this method had so far a number of unaccounted factors, the effect of which could not be correctly evaluated. Therefore, the accuracy of this method was practically unknown. To establish a quantitatively accurate calorimetric measurement, we can either model all processes occurring during the experiment[23] or as shown in this paper, the method can be calibrated with any well-known reaction. In this paper, we demonstrate that the calibration of the calorimetric technique allows a precise measurement of the reaction energy between almost any pair of reactants. For the calibration, we will use the reaction between C atoms and $H_2$. Then this new method is applied to the reactions of C atoms with $C_2H_2$ and $O_2$ molecules to demonstrate its potential.

**Calorimetry measurements**

Figure 1 shows the principal scheme of reactions occurring inside He droplets. Reactants are incorporated sequentially into He droplets. In less than 1 μs they meet inside the droplets[24]. Exothermic reactions with zero energy barriers in the entrance channel proceed spontaneously upon encounters. The reaction energy is transferred to the droplets leading to a size reduction. Each evaporated He atom removes about 7.2 K of energy[13]. By measuring the size of He droplets before and after the evaporation, we can calculate the amount of energy released in the reaction.

  We tested two different approaches for measuring the He droplets sizes. The pressure method (PM) was described previously[17, 18] and is based on measuring the He pressure in the detector chamber. As the He is evacuated at a constant speed from the vacuum chamber, the measured pressure is directly proportional to the number of He atoms arriving in the chamber. In the ion signal method (IOS) the evaluation of He droplet sizes is based on the probability of the collisions between He droplets and electrons. As it is shown in the method section this method needs further development but is very promising.

The depletion values measured by the PM method at different doping efficiencies of He droplets with H2 are provided in Figure 2, displaying a linear dependence of the depletion values on the doping efficiency. This is in line with the theoretical expectation. The depletion D of the He droplet beam caused by picking up C atoms is given by

$$D = D_{\text{pk}} + P_{R_1} P_{R_2} \frac{E_R}{5N_{He}}, \tag{1}$$

where $D_{\text{pk}}$ is the depletion caused by the pick up of C atoms by He droplets without a second reactant, $P_{R_1}$ and $P_{R_2}$ are the fractional abundances of He droplets doped with the first and second reactant, respectively, $N_{He}$ is the initial number of He atoms per droplet, and $E_R$ is the reaction energy given in reciprocal centimeters. As $D_{\text{pk}}$ is independent of the amount of the second reactant, this quantity has been extracted from the measured depletion values (see SI for the C + H2 data where $D_{\text{pk}}$ is not extracted).

In our experiment, we only vary the doping efficiency of He droplets with a second reactant $P_{R_2}$. Therefore, we expect a linear dependence of the measured depletion on the $P_{R_2}$ value. This behavior is indeed seen for all three considered reactions (see Fig,2).

We can now linearly fit the experimental data and obtain the following slope after Equation 1:

$$m = P_{R_1} \frac{E_R}{5N_{He}}. \tag{2}$$

We can now derive the reaction energy $E_R$ after the other quantities in Equation (2) have been determined. The fraction of He droplets doped with C atoms can be approximately evaluated based on the mass spectrum of doped helium droplets. Ionization of the helium droplets doped with a single C atom results in the formation of $He_n^+$, $C^+$, and $CHe_n^+$ cations. Ionization of He droplets doped with two C atoms produces $He_n^+$ and $C_2^+$ cations. Therefore, the ratio of the intensities of ion peaks $^{13}C + ^{13}CHe_n$ and $^{13}C_2$ should show the ratio of the number of He droplets doped with one and two carbon atoms. The probability of pick-up of k species follows a Poisson distribution[13]. Based on this distribution, we calculated the doping efficiency of He droplets with C atoms $P_{R_1}$ to be 6.5%. We intentionally worked at low doping conditions to minimize the contribution of the droplets containing more than a single C atom. Assuming that the initial He droplet size is equal to 15000 He atoms[13] and taking into account the obtained slope of the linear fit from Figure 2, we obtained that $E_R = 49782$ K. The obtained value is significantly larger than 38623 K which is derived based on atomization energies of $CH_2$ and $H_2$[25, 26]. This is not surprising, considering uncertainties in the determination of $P_{R_1}$. Our evaluation of this value is affected by the possibility that $C_2$ molecules could dissociate upon ionization and that formation of the $C_2$ molecule reduces the ionization cross-section of the He droplet. Both effects reduce the intensity of ion signals measured on the mass of $C_2$ molecules, leading to an underestimation of the doping efficiency of He droplets with C atoms. If we perform the reverse calculation, using the correct value of the reaction energy, we obtain $P_{R_1} = 8.4$% instead of 6.5%. Therefore, even a small uncertainty in estimating the doping efficiency of He droplets generates a relatively large error in the measured reaction energy. This behavior provides the main contribution to the uncertainties of the method. However, the other parameters in Equation 2 can also add to an error in the measurement. The initial size of He droplets is based on measurements with different experimental setups. A slight variation in the size of He droplets produced by different experimental setups can certainly be expected. Additionally, we cannot monitor a single He droplet, but rather a population of droplets. The log-normal size distribution of the He droplets with a logarithmic tail towards larger sizes results in an error when using an average size of He droplets. Moreover, in a range of very small droplet sizes, the binding energy of He atoms to a droplet is decreasing with smaller sizes of He droplets. Therefore, it becomes less than 7.2 K, which is considered in Equation 2.

Because of these uncertainties, a much better approach is to combine all these (uncertain) parameters in a single quantity. The determination of this quantity would calibrate our calorimetric technique, allowing us to obtain precise data. Moreover, such a calibration can take into account possible effects that have not yet been considered. Here the similarity to a classical calorimeter has to be mentioned, where the calibration procedure is absolutely necessary for precise measurements. With the corresponding change in Equation 2, we can redefine the reaction energy:

$$E_R = K_{cal} m \qquad (3)$$

where, $K_{cal}$ is the calibration constant. The reaction of C atoms with $H_2$ is nicely suited for the calibration. This low-temperature reaction has only two possible outcomes. If an energy barrier in the reaction pathway would be present, the formation of the weakly bound $CH_2$ complex would lead to the release of only a small amount of energy ~863 K[27], similar to the reaction of C atoms with Ar, where the formation of an only weakly bound van der Waals complex takes place. This is not consistent with the experimental results as the C + $H_2$ and C + Ar slopes are clearly different (see Fig.2a). The only other possibility is the formation of a triplet state of the HCH molecule with the release of 38623 K of energy, which is derived based on atomization energies of $CH_2$ and $H_2$[25, 26]. Therefore, using these data, we can determine the calibration constant and obtain a value of 1170379 K.

**Results**

We now apply this calibration technique to measure the reaction energy released in the $C_2H_2$ + C reaction. As seen in Figure 2, the measured slope for this reaction is 0.023. Therefore, we obtain a released energy of 26918 K. The expected error is below 1300 K. The deviation of the measured depletion values from the linear fit was found to be mainly associated with the instability of the flux of C atoms. Therefore, a further increase in the accuracy can be achieved either by enhancement of the stability of the atomic carbon source or by simply increasing the number of data points used for the linear fit.

This reaction has been previously studied in the gas phase by crossed molecular beam techniques and by quantum chemical computations[28,29]. In the gas phase, the reaction C +

$C_2H_2 \rightarrow C_3H + H$ takes place, which is followed by the release of 963 K of energy. The other possible reaction products predicted by computations are listed in Table 1.

As can be seen, a very close match between the amount of energy release derived in our experiment (26918 K) and that predicted in the highest level of computation with the claimed accuracy of about 360 K, is achieved for the formations of the triplet cyclic-$C_3H_2$ product (25918 K). Therefore, we can confidently conclude that this product is formed in our experiment.

This is in line with theoretical predictions. The formation of dehydrogenated products, as observed for this reaction in the gas phase, is not expected when the third body is available for the dissipation of reaction energy. Additionally, a larger number of linear products was detected in cross-molecular beam experiments going to higher collision energies of the reactant[29]. Therefore, the formation of all linear products is predicted to be associated with an energy barrier in the reaction pathway and is not expected at low temperatures.

The other studied reaction $C + O_2$, has been previously investigated by a conventional method on the surface of Pt crystal, where the initial formation of CO and O was found[30]. This reaction is highly exothermic with an energy release of 69559 K. Even a larger amount of energy release of 132839 K is expected for the formation of the $CO_2$ product[31]. As can be seen from Figure 2b,c, for each measurement its own calibration curve was obtained. Using the same calculations as before, we obtained similar values of 62726 K and 66487 K in the experiments with smaller and larger helium droplets, correspondingly. This indicates the formation of CO and O products, which is in line with the previous studies[30]. The large helium droplets were used to observe the secondary reaction CO + O and the formation of $CO_2$, which was not successful. This could be due to the presence of an energy barrier in the CO + O reaction pathway[32]. However, due to the presence of several states of $CO_2$ molecule energy levels which are only known with a low accuracy[33, 34], we cannot absolutely exclude the formation of $CO_2$ molecules in the excited electronic state. Additionally, these measurements have a larger error due to the use of larger He droplets.

**Application**

The developed calorimetric technique provides excellent accuracy in determining the reaction energies, which is sufficient for comparison with quantum chemical computations and allows us to identify the reaction channel in studied reactions. Additionally, results obtained with the developed method can be used to get more information on the pathways of gas-phase reactions, as products of surface reactions are often intermediate states of the corresponding gas-phase reactions. For example, in the $C + C_2H_2$ reaction, we do not observe a change in multiplicity with the formation of the most favorable singlet state of c-$C_3H_2$ (see Table 1). This means that the time needed for such a transition is considerably longer than the time scale of the experiment, which is about a few milliseconds. This also shows that the reaction channel that includes this singlet-triplet crossing is not active in the gas phase in contrast to the suggestion of early quantum chemical computations[28].

The developed method provides data, which can be directly compared with the energy level diagram provided by quantum chemical computations. From this comparison, one can estimate the presence of energy barriers in a reaction pathway and find active reaction channels of reactions occurring in the gas phase or on surfaces with low catalytic activity. Due to the presence of the third body in our experiment, the final product of the reaction is the same as it would be formed in the reaction on a chemically inert surface. This provides guidance, which reaction could operate on astrophysically relevant surfaces, for example, water ice surfaces. This is a good approximation, as the catalytic activity of water ice surfaces was found to be rather small and in most cases led to the reduction of the energy barrier of the reactions[35-37]. Therefore, water ice is not expected to provide large effects on the barrierless reactions studied by our method.

However, for the highest precision of chemical models, the catalytic activity of the real surfaces needs to be taken into account. The catalytic activity can be understood as a modification of levels in an energy level diagram of a reaction. Therefore, the comparison of the original unmodified and the catalytically modified diagrams are expected to provide fundamental information, which will help to understand the catalytic mechanisms, and therefore, how the chemical reactivity may change on different types of surfaces. This is especially important for the case of astrochemistry, as the knowledge of real grain surfaces present in the ISM is rather limited. Both energy level diagrams can be obtained in the computations. We have demonstrated that our method is nicely suited for comparison with the unmodified energy level diagram obtained in the computation. Our method can be extended to study the catalytic effects of chemically active surfaces. For this purpose, the He droplets can be pre-doped by clusters modeling grain surfaces. The results of such measurement can then be compared with the modified energy level diagram.

Besides this, the method could also be used to predict rates of Eley-Rideal reactions. The fast reaction at $T$ = 0.37 K shows zero energy barrier in the reaction pathway; therefore the reaction rates will be largely defined by collision rates. At the same time, the rates of Langmuir-Hinshelwood reactions also depend on diffusion rates of reactants on surfaces, which are not defined by our method.

**Astrophysical implications**

Already the first reactions studied in this paper provide important results on the surface chemistry of carbon atoms. In the interstellar medium, carbon atoms are accreted by dust grains with ice mantles and should interact with $H_2$ molecules. In the astrochemical reactions databases, the barrier for this reaction is 2500 K[8], which makes the reaction impossible at low temperatures. Therefore, the observed rapid low-temperature C + $H_2$ surface reaction will dramatically change the complete carbon chemistry network. The reaction C + $C_2H_2$ → c-$C_3H_2$ is also barrierless, but is not present at all in astrochemical reaction databases. At the same time, the formation of cyclic products in this surface reaction could explain the observed increase in $N$(cyclic-$C_3H_2$)/$N$(linear-$C_3H_2$) ratio going to denser regions of the ISM. The ratio increases from 3 in the diffuse ISM up to 100 in dark clouds[38]. The product of the C + $O_2$ reaction is the same as in the gas phase. Therefore, this reaction should provide less influence on the chemistry of the ISM.

The transition of molecular clouds from atomic gas starts with carbon present in atomic form. Deeper in the molecular cloud, atomic carbon is converted to CO, but is also accreted on the dust contributing to the formation of molecules. At the surface of molecular clouds, C atoms are efficiently formed due to the photodissociation of CO molecules, and C atoms are even produced inside the dense part of clouds due to Lyman α photons[39, 40]. The formed C atoms again accrete on the grain surfaces. This demonstrates that reactions with carbon atoms are potentially of great importance for the formation of COMs in astrophysical environments. Our method can be applied to study the reactions between any species, with the only known exception of hydrogen atoms, which cannot be picked up by helium nanodroplets. Therefore, we expect the method to have great potential to study the formation channels of large COMs.

**Conclusions**

In this paper, we established a new technique for the precise measurement of reaction energies, based on the He droplet approach. For most of the reactions, the achieved accuracy using this approach will be higher than achieved in the highest level of quantum chemical computations. The comparison of the experimental data with quantum chemical calculations provides fundamental information on reaction mechanisms as well as potential energy surfaces of reactions. This allows the prediction of many aspects of astrochemical reactions. Due to the presence of the third body in the experiment, the obtained results are much more suitable for the prediction of the outcome of the astrochemical surface reactions compared to the results of the gas-phase studies. The found barrierless reactions of C with $H_2$ and $C_2H_2$ are expected to have a large impact on the chemistry of dense regions of the ISM.

**Methods**

**The helium droplet setup.** The He droplet apparatus is basically the same as reported in earlier publications[17, 41]. Supplementary Figure 1 shows the principal scheme of the setup. He droplets are produced in the continuous expansion of the compressed He (99.9999% purity) through a cooled 5 µm nozzle with a fixed helium pressure ($p$ = 20 Bar). The size of helium droplets was varied by changing the temperature of the nozzle[13]. The produced helium droplets, after passing the skimmer ($\varnothing$ 0.3 mm), arrive in the main 300 mm long chamber where two different pick-up cells are installed. The pressure in this chamber is monitored by the ion gauge (Leybold Heraeus IE-20). The He droplets are doped with $^{13}C$ atoms using a dedicated atomic C source[42]. The $^{13}C$ isotope is used to simplify a mass spectrometric detection of the products of chemical reactions. was installed. The gases, $H_2$ (Air Liquide, 99.9995% purity), oxygen (Air Liquide, 99.999% purity), argon (Air Liquide, 99.999% purity), and $C_2H_2$ (Air Liquide, unspecified purity), were provided to a 40 mm long pick-up cell equipped with an ion gauge (Leybold Heraeus IE-20) from outside through a leak valve. The purity of all reactants was additionally controlled by recording the mass spectra of doped helium droplets. The ion gauges in the main 300 mm long chamber as well as in the pick-up 40 mm long cell, are used to compute the doping efficiency of He droplets by gases based on the pressure rise upon opening the leak valve. The pressure rise in the main chamber is about two orders of magnitude lower compared to the pressure rise in the 40 mm pick-up cell. Therefore, gas-phase reactions cannot be significant. The pressure of the residual gas in the reaction chamber was about $10^{-7}$ mbar, which is extremely high compared to the condition ($10^{-11} - 10^{-12}$ mbar) required for classical studies of surface reactions. The extremely small lifetime of He droplets of about few milliseconds and the small size of the nanodroplets considerably reduce the requirement for the number of residual species in a vacuum chamber. A negligibly small doping of helium nanodroplets with residual species is already achieved when the pressure of residual gases is in the $10^{-7}$ mbar range, considerably relaxing the requirement for the vacuum conditions.

To estimate the doping efficiency of He droplets with C atoms is more challenging, as the number density of C atoms in the gas phase cannot be simply evaluated. Therefore, for an approximate evaluation, the values were obtained based on the ratio of the ion signals measured on the masses of $^{13}C + ^{13}CHe_n$ and $^{13}C_2$. The stability of the doping conditions for C atoms was continuously monitored during the experiment based on the ratio of $He_3$ and $^{13}C$ peaks. After passing two doping regions, the He droplet goes through an 800 µm aperture and arrives at the pre-pumping vacuum chamber evacuated by the turbo pump. This chamber is used to lower the helium gas load to the next UHV chamber, which is evacuated by two ion pumps. The pressure in the detector chamber was recorded with the precision ion gauge (Varian UHV-24). The pressure values from all ion gauges were monitored with 1 kHz frequency. Averaging over 1000 data points resulted in one data point per second. A quadruple mass spectrometer (QMS) is installed in the detector chamber and is used in combination with the electron impact ionizer. Electrons with energies of 70 eV were employed for ionization.

**Measurement of the depletion of the He droplet beam caused by incorporation of C atoms by the PM method.** To measure the relative size change of the He droplets after the reaction by the PM, the following procedure was established. The pressure in the detector chamber during the depletion measurement is indicated in Supplementary Figure 2. The figure also shows the current passing through the filament of the C source. As can be seen, the rise of the current up to the operational value causes a decrease in the pressure in the detector chamber. This happens due to the reduction of the size of helium droplets after the pick-up of

C atoms. However, only the He droplets, with C atoms contribute to the total depletion of the helium droplet beam intensity. To measure the depletion of the He droplet beam, the atomic C source was switched on and off every 30 s. The 20 s gates were used to average the pressure values for both states of the atomic C source. The gate, which was used to get the values during the operational state of the C atom source, started 8 s after the current rose. This provides time for stabilizing the pressure in the detector chamber. Similarly, the pressure of the residual gas in the detector chamber was obtained 10 s after the complete blocking of the He droplet beam by a shutter in the source chamber. The total depletion of the helium droplet beam was calculated by $(P_{\text{OFF}} - P_{\text{ON}})/(P_{\text{OFF}} - P_{\text{RES}})$, where $P_{\text{OFF}}$ and $P_{\text{ON}}$ are the pressures during the time, when the atomic C source is switched off and on, respectively, $P_{\text{RES}}$ is the residual pressure in the detector chamber.

**Measurement of the He droplet sizes by the IOS method.** We crossed the He droplet beam with a pulsed electron beam. To do so, an electron impact ionizer, identical to the one which is used in our QMS, was installed in the main chamber. The ionizer has two 4 mm diameter openings so that the He droplet beam passes it undisturbed. Upon the collision of He droplets with electrons, the droplets are either destroyed or deflected. This causes a depletion peak in the ion signal detected by QMS tuned on the masses of the He dimer, which can be seen in Supplementary Figure 3a. The intensity of depletion peaks is indeed proportional to the geometrical cross-section of the droplets. The probability of a collision between an electron and a droplet is given by equation $P = 1 - \exp(-n_{\text{el}}\sigma_{dr}l_{\text{ion}})$, with $\sigma_{dr}$ being the geometrical cross-section of the droplet, $n_{\text{el}}$ is the number density of electrons, and $l_{\text{ion}}$ is the length of the ionization volume. As can be seen from Supplementary Figure 3b, the measured depletion is not directly proportional to the expected collision probability. This indicates a more complex behavior after the collision of the electron with the droplet. Moreover, due to the short duty cycle, this method was found to be less precise compared to the PM. Therefore, we selected the PM to measure the relative size change of the He droplets, and the IOS was not used. However, the IOS has a strong potential as it can be used to measure the reduction in the size of the doped He droplets exclusively, while the PM applies to all droplets. IOS would have a strong advantage if a reaction has two or more pathways resulting in different products. A disadvantage of the PM is the fact that we measure only an average value of the reaction energy. If the reaction proceeds via several competitive channels, the method in the current realization could not discriminate between different channels. In the future, the application of the IOS method for measuring the sizes of He droplets as it is shown in Supplementary Figure 3 allows for solving this problem. In this case, tuning the mass spectrometer on the mass of a specific reaction product will allow us to measure the energy released in the reaction pathway associated with an exact product. Additionally, the measured reaction energies will not depend on the estimation of the fraction of He droplets doped with reactants, as only doped droplets will be probed.

**The advantages of using the slope for the measurement of reaction energies.** The use of the slope of the linear fit for the measurement of reaction energies has several advantages. First, the accuracy of the measurement is considerably increased due to the evaluation of several data points at different doping conditions. However, even more importantly the method is robust to fluctuations of the depletion caused by the incorporation of a single reactant. We will illustrate this with an example: Supplementary Figure 4 shows two fits to the reactions of C atoms with $H_2$. One data set was recorded when a less pure carbon powder was loaded in the C source. As a result, the source produces additional CO molecules, which were also picked up by He droplets. The energy released in the reaction between CO molecules and C atoms resulted in an enhanced depletion. The evaluation of absolute values of the depletion would lead to a considerable error in the measurement of the reaction energy. Contrary, the use of the slope, as proposed here, almost entirely removes this problem. Independently of the originally higher depletion value, the measured slope is the same as that obtained with a pure atomic C source. This new method allows us to measure the depletion correctly and forms the basis of a much improved measurement of reaction energies on inert surfaces.

**The experiments at high doping conditions.** The linear dependences shown in Figure 2 were found to continue in the range of high doping efficiencies. At these conditions, calculations based on Poisson distribution predict that helium droplets pick up several molecules of the second reactant. Therefore, a product of the first reaction has a chance to react with molecules, which are picked up later by the same droplet. This effect cannot influence the reactions of C atoms with $H_2$ and $O_2$ molecules, as HCH and CO are not reactive towards $H_2$ and $O_2$, respectively. The product of the C + $C_2H_2$ reaction is potentially reactive towards additionally picked-up $C_2H_2$ molecules. However, the experiments with $C_2H_2$ were performed in the configuration when C atoms are incorporated in the first position, and the time between incorporations of $C_2H_2$ molecules is considerably larger than the time needed for reactants to meet inside a droplet. As a result, the pick-up cross-section of the helium droplets is considerably reduced after the first reaction, and the probability to pick up additional $C_2H_2$ molecules becomes very small. Therefore, we expect that under the conditions used for our measurements, the influence of these additional reactions is very slight. This is confirmed by the continuation of the linear dependences shown in Figure 2 in the high doping range. However, generally, the possibility of such an effect should be kept in mind. It is expected to be more noticeable in the case of larger helium droplets and can be recognized due to the deviation from the linear dependence towards higher depletions in a high doping range.

**Data availability.** The data that support the plots within this paper and other findings of this study are available from the corresponding author upon reasonable request.

**Acknowledgements**
The authors are grateful for the support of the Max Planck Society and the DFG (contract No. KR 3995/3-1).

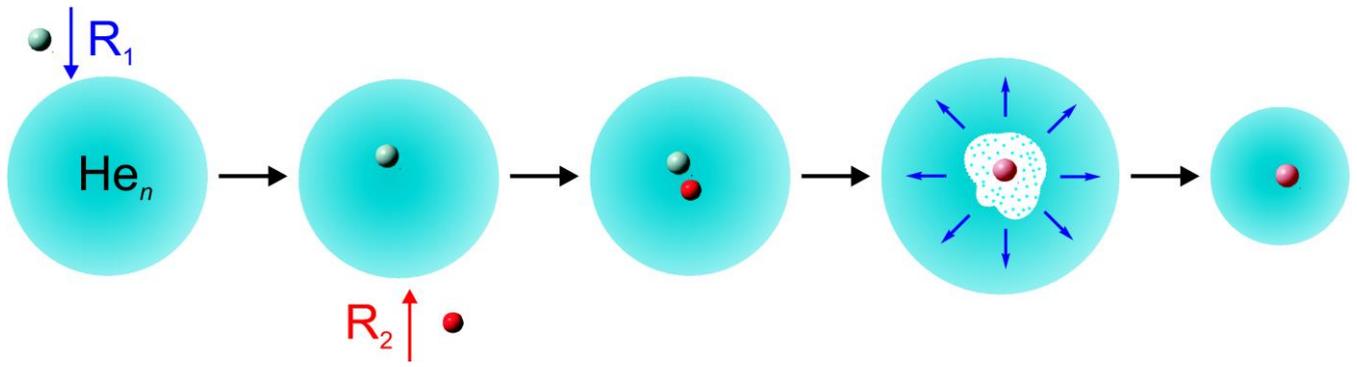

**Figure 1.** Schematic representation of the chemical reactions occurring inside He droplets. $R_1$ and $R_2$ are the two different reactants.

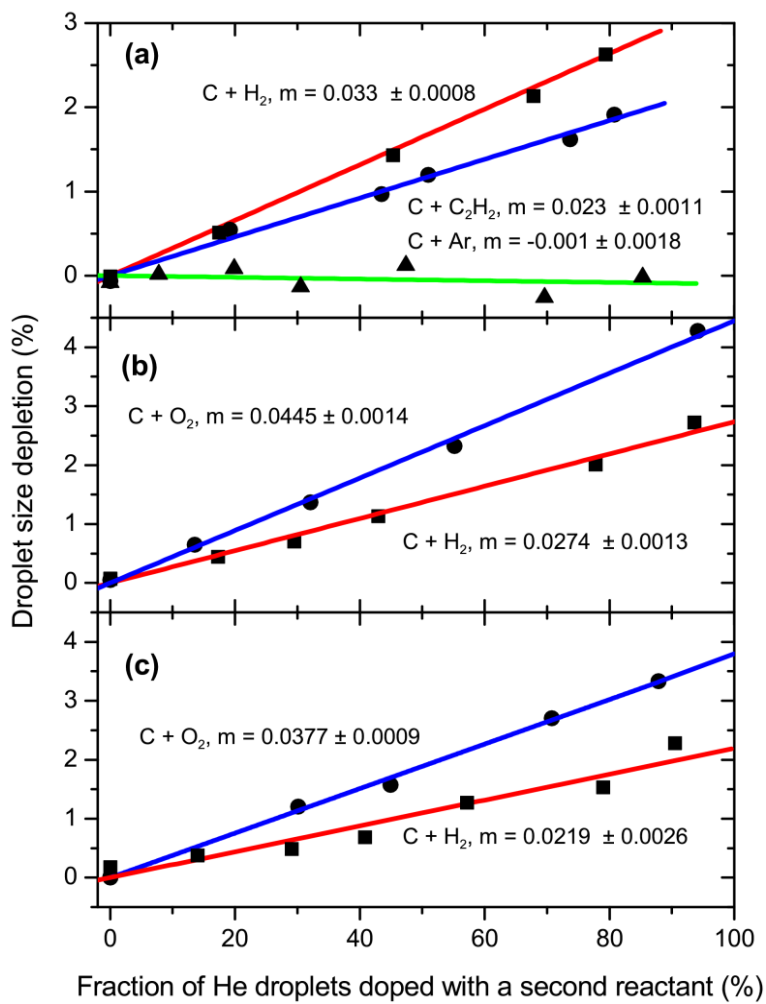

**Figure 2.** The measured depletion of the He droplet beam that is caused by exothermic reactions of a fixed amount of C atoms as a function of the efficiency of doping of the He droplets with a second reactant. The type of the second reactant is given in the reaction specified for each data set. The initial helium droplet sizes were 15000 (a), 20500 (b), and 145000 He atoms (c). The curves for the calibration reaction C + $H_2$ are shown in all diagrams. The errors of the slope measurements are defined based on the standard deviations of the measured depletion data points from the theoretical linear function.

**Table 1.** Energy levels of the C + $C_2H_2$ reaction obtained from quantum chemical computations using (a) UCCSD(T)/QZ2P (Ref. 29) and (b) MP4/6-311G(d,p) (Ref.28) levels of theory. Energies are given in units of kelvin. Electronic states given in parentheses are taken when possible from [28].

| State | Energy[a] | Energy[b] |
|---|---|---|
| C ($^3P_0$) + $C_2H_2$ | 0 | 0 |
| CCHCH ($^3A''$) | -16488 | -11261 |
| c-$C_3H_2$ ($^3A$) | -25918 | -24275 |
| c-$C_3H_2$ ($^1A_1$) | | -50664 |
| HCCCH ($^3A''$) | -46652 | -39130 |
| HCCCH ($^1A$) | | -37076 |
| $CCCH_2$ ($^3B_1$) | -30429 | |
| c-$C_3H$ ($^2B_1$) + H | -963 | -3280 |
| ℓ-$C_3H$ ($^2\Pi$) + H | -240 | -8343 |

# Supplementary Information

**Measurement of the He droplet sizes by the IOS method**
We crossed the He droplet beam with a pulsed electron beam. To do so, an electron impact ionizer, identical to the one which is used in our QMS, was installed in the main chamber. The ionizer has two 4 mm diameter openings so that the He droplet beam passes it undisturbed. Upon the collision of He droplets with electrons, the droplets are either destroyed or deflected. This causes a depletion peak in the ion signal detected by QMS tuned on the masses of the He dimer, which can be seen in Supplementary Figure 3a. The intensity of depletion peaks is indeed proportional to the geometrical cross-section of the droplets. The probability of a collision between an electron and a droplet is given by equation $P = 1 - \exp(-n_{\text{el}}\sigma_{dr}l_{\text{ion}})$, with $\sigma_{dr}$ being the geometrical cross-section of the droplet, $n_{\text{el}}$ is the number density of electrons, and $l_{\text{ion}}$ is the length of the ionization volume. As can be seen from Supplementary Figure 3b, the measured depletion is not directly proportional to the expected collision probability. This indicates a more complex behavior after the collision of the electron with the droplet. Moreover, due to the short duty cycle, this method was found to be less precise compared to the PM. Therefore, we selected the PM to measure the relative size change of the He droplets, and the IOS was not used. However, the IOS has a strong potential as it can be used to measure the reduction in the size of the doped He droplets exclusively, while the PM applies to all droplets. IOS would have a strong advantage if a reaction has two or more pathways resulting in different products. A disadvantage of the PM is the fact that we measure only an average value of the reaction energy. If the reaction proceeds via several competitive channels, the method in the current realization could not discriminate between different channels. In the future, the application of the IOS method for measuring the sizes of He droplets as it is shown in Supplementary Figure 3 allows for solving this problem. In this case, tuning the mass spectrometer on the mass of a specific reaction product will allow us to measure the energy released in the reaction pathway associated with an exact product. Additionally, the measured reaction energies will not depend on the estimation of the fraction of He droplets doped with reactants, as only doped droplets will be probed.

**The advantages of using the slope for the measurement of reaction energies**
The use of the slope of the linear fit for the measurement of reaction energies has several advantages. First, the accuracy of the measurement is considerably increased due to the evaluation of several data points at different doping conditions. However, even more importantly the method is robust to fluctuations of the depletion caused by the incorporation of a single reactant. We will illustrate this with an example: Supplementary Figure 4 shows two fits to the reactions of C atoms with $H_2$. One data set was recorded when a less pure carbon powder was loaded in the C source. As a result, the source produces additional CO molecules, which were also picked up by He droplets. The energy released in the reaction between CO molecules and C atoms resulted in an enhanced depletion. The evaluation of absolute values of the depletion would lead to a considerable error in the measurement of the reaction energy. Contrary, the use of the slope, as proposed here, almost entirely removes this problem. Independently of the originally higher depletion value, the measured slope is the same as that obtained with a pure atomic C source. This new method allows us to measure the depletion correctly and forms the basis of a much improved measurement of reaction energies on inert surfaces.

**The experiments at high doping conditions**
The linear dependences shown in Figure 2 were found to continue in the range of high doping efficiencies. At these conditions, calculations based on Poisson distribution predict that helium droplets pick up several molecules of the second reactant. Therefore, a product of the first reaction has a chance to react with molecules, which are picked up later by the same droplet. This effect cannot influence the reactions of C atoms with $H_2$ and $O_2$ molecules, as HCH and CO are not reactive towards $H_2$ and $O_2$, respectively. The product of the C + $C_2H_2$ reaction is potentially reactive towards additionally picked-up $C_2H_2$ molecules. However, the experiments with $C_2H_2$ were performed in the configuration when C atoms are incorporated in the first position, and the time between incorporations of $C_2H_2$ molecules is considerably larger than the time needed for reactants to meet inside a droplet. As a result, the pick-up cross-section of the helium droplets is considerably reduced after the first reaction, and the probability to pick up additional $C_2H_2$ molecules becomes very small. Therefore, we expect that under the conditions used for our measurements, the influence of these additional reactions is very slight. This is confirmed by the continuation of the linear dependences shown in Figure 2 in the high doping range. However, generally, the possibility of such an effect should be kept in mind. It is expected to be more noticeable in the case of larger helium droplets and can be recognized due to the deviation from the linear dependence towards higher depletions in a high doping range.

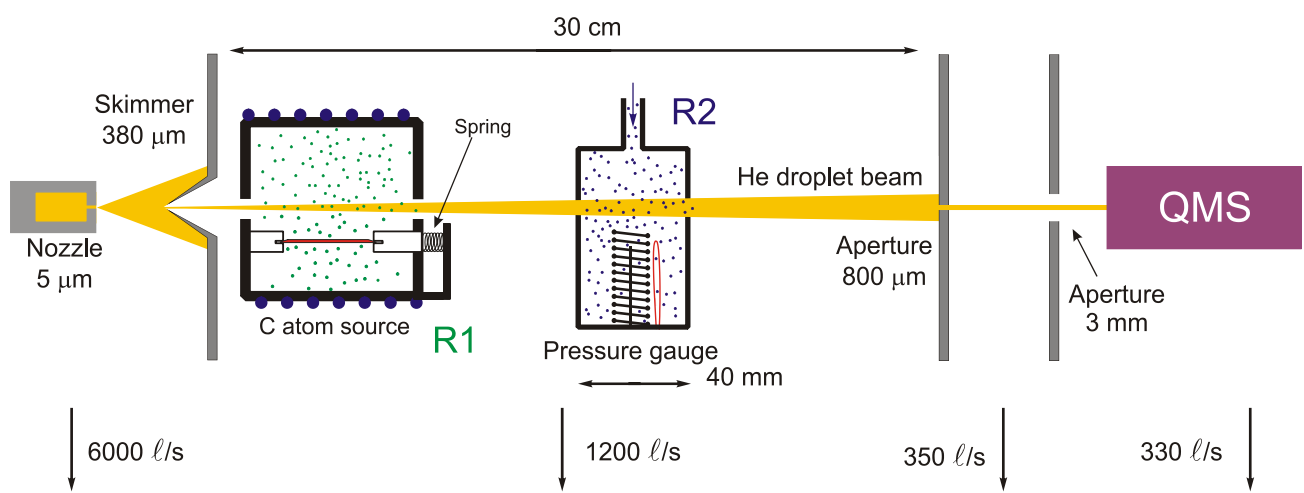

**Supplementary Figure 1.** The principal scheme of the experimental setup used for the measurement of the reaction energies released in the reaction C + $C_2H_2$, occurring inside superfluid helium nanodroplets. To study the reaction of C + $O_2$ the oxygen was incorporated before the carbon atoms.

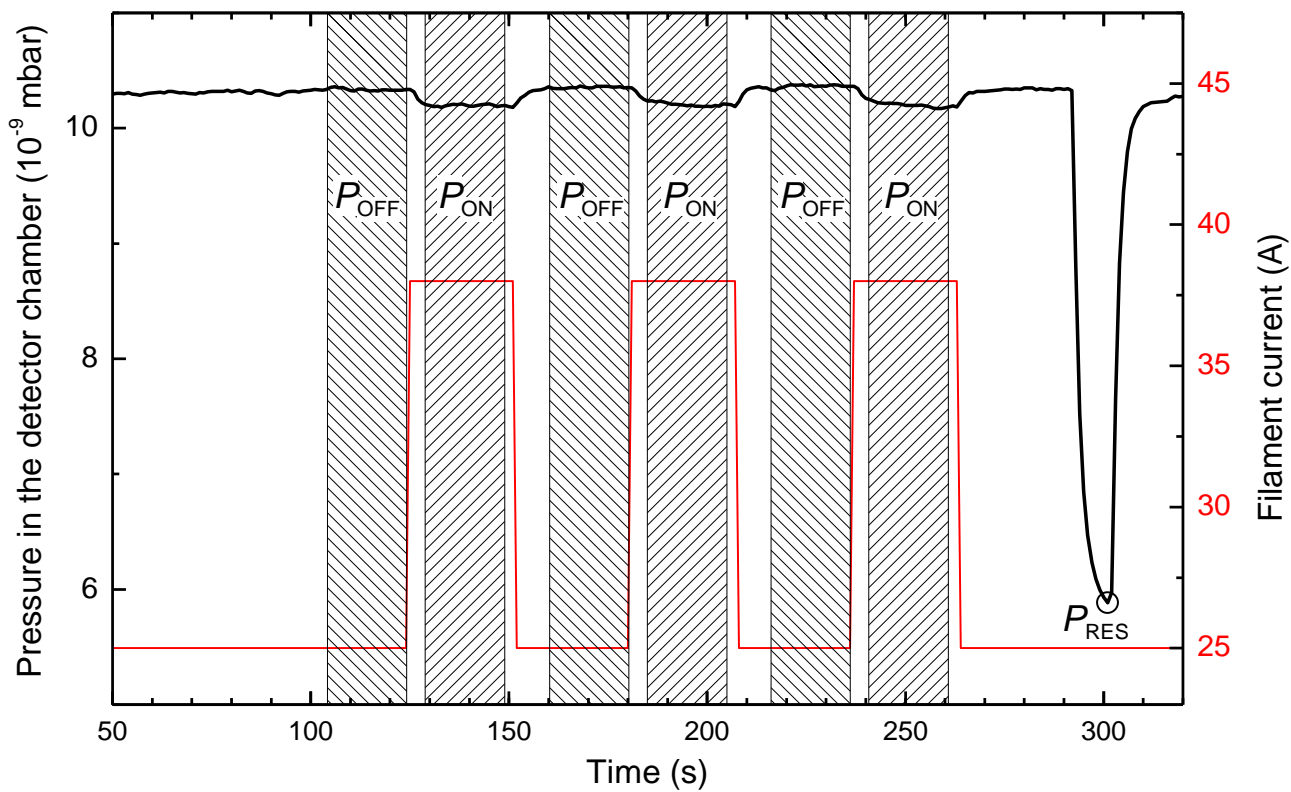

**Supplementary Figure 2.** Pressure in the detector chamber and the current passing through the filament of the atomic C source during the depletion measurement. The rectangular hatched areas show times when pressure was measured to calculate the depletion. The circle marks the value used as a residual pressure ($P_{RES}$) in the detector chamber. In this experiment, 80.8% of all helium droplets were doped with $H_2$ molecules.

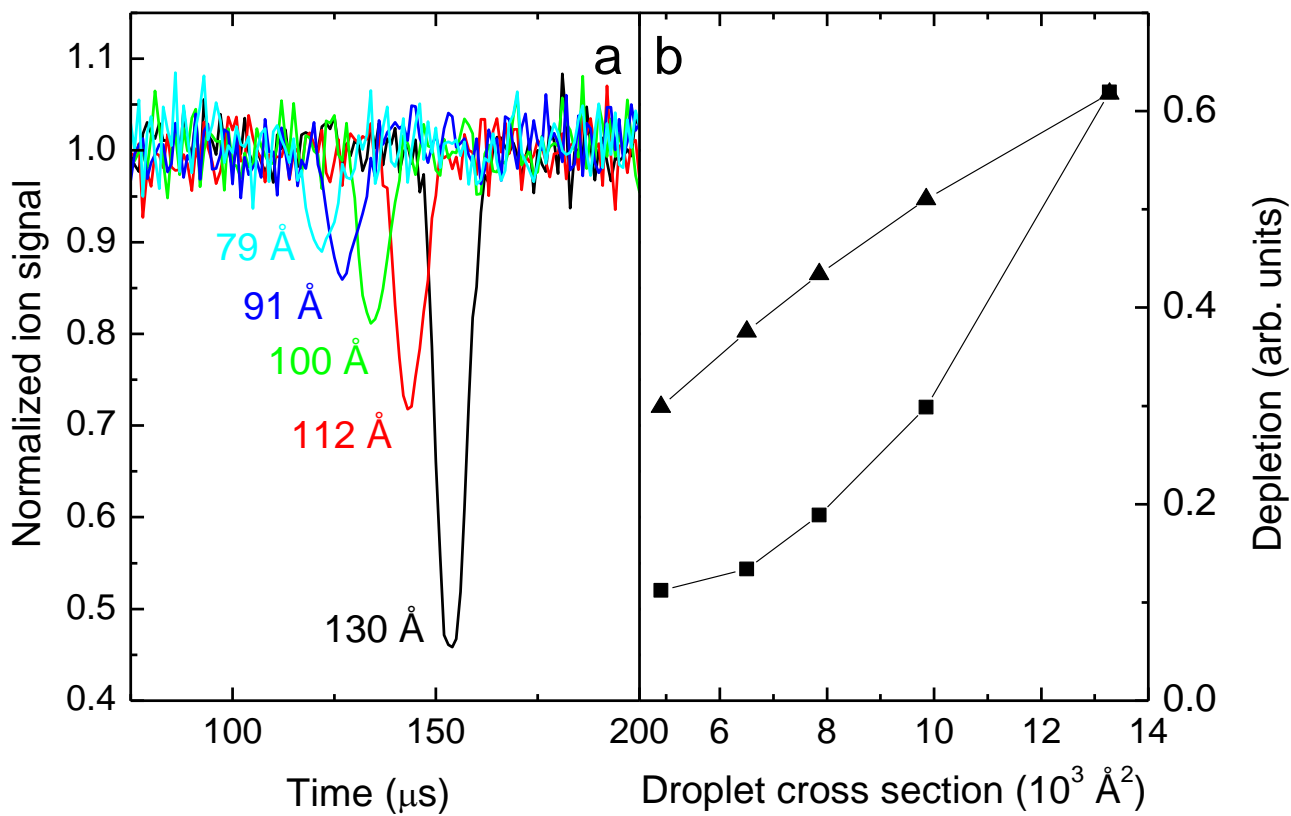

**Supplementary Figure 3.** Ion signals on the mass of the He dimer measured by QMS as a function of time (a). The depletion peaks are due to the interaction of the He droplet beam with the pulsed electron (100 eV) beam. The numbers show the diameter of the used He droplets. Panel (b) shows the dependence of the measured area of the observed depletion peaks in panel (a) on the geometrical cross-section of the droplet (square data points), while triangles show the calculated probabilities of collision between He droplets and electrons.

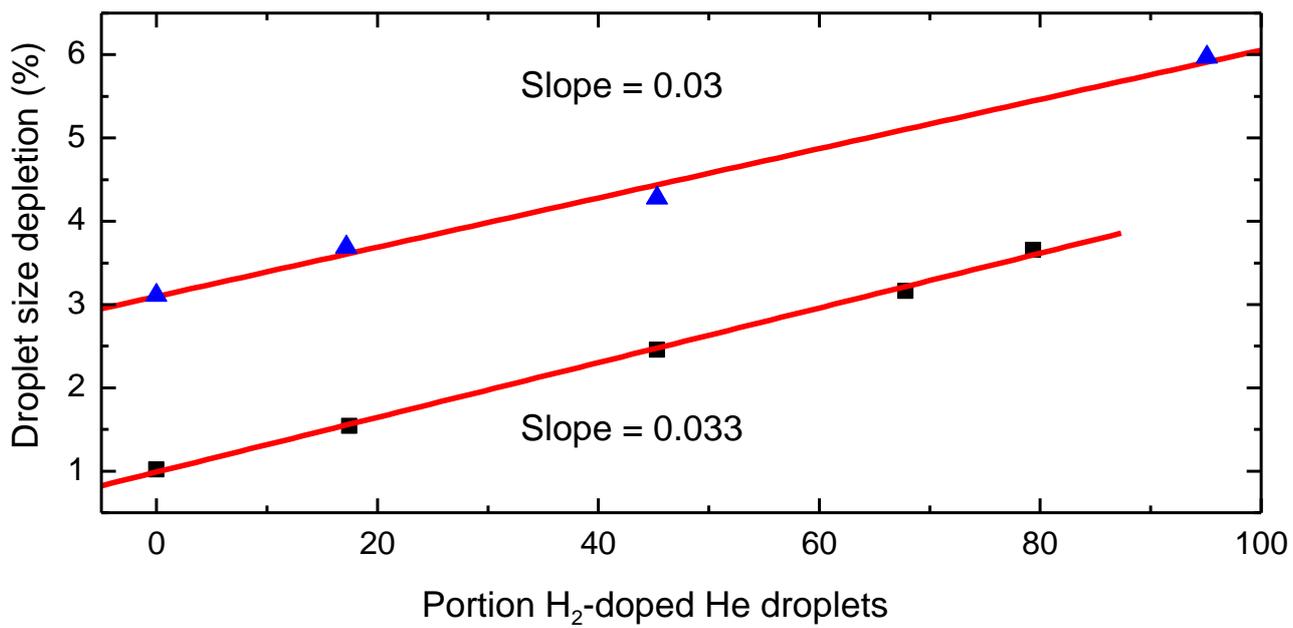

**Supplementary Figure 4.** The measured depletion of the He droplet beam was caused by the incorporation of a fixed amount of C atoms as a function of the efficiency of doping of the He droplets with hydrogen at a later stage. Triangles show the experimental data obtained under the presence of impurities causing additional depletion compared to the clean experiment shown by square data points.